\documentclass[english,showpacs,preprint,superscriptaddress]{revtex4-1}
\usepackage[T1]{fontenc}
\usepackage{babel}
\usepackage{mathtools}
\usepackage{amsmath}
\usepackage{amsthm}
\usepackage{amssymb}
\usepackage[unicode=true,
 bookmarks=false,
 breaklinks=false,pdfborder={0 0 1},backref=false,colorlinks=false]
 {hyperref}

\makeatletter

\usepackage{amsthm}\usepackage{latexsym}\usepackage{bm}\usepackage{amsfonts}

\allowdisplaybreaks
\usepackage{babel}
\usepackage{babel}

\usepackage{babel}

\makeatother

\begin{document}
\title{High-order Field Theory and Weak Euler-Lagrange-Barut Equation for
Classical Relativistic Particle-Field Systems}
\author{Peifeng Fan}
\email{corresponding author: pffan@szu.edu.cn}

\affiliation{Key Laboratory of Optoelectronic Devices and Systems, College of Physics
and Optoelectronic Engineering, Shenzhen University, Shenzhen 518060,
China}
\affiliation{Advanced Energy Research Center, Shenzhen University, Shenzhen 518060,
China}
\author{Qiang Chen}
\email{corresponding author: qiangchen@zzu.edu.cn}

\affiliation{National Supercomputing Center in Zhengzhou, Zhengzhou University,
Zhengzhou, Henan 450001, China}
\author{Jianyuan Xiao}
\email{corresponding author: xiaojy@ustc.edu.cn}

\affiliation{School of nuclear science and technology, University of Science and
Technology of China, Hefei, Anhui 230026, China}
\begin{abstract}
It is widely accepted that conservation laws, especially energy-momentum
conservation, have fundamental importance for both classical and quantum
systems in physics. A widely used method to derive the conservation
laws is based on Noether's theorem. However, for classical relativistic
particle-field systems, this process is still impeded. Different from
the quantum situation, the obstruction emerged when we regard the
particle's field as a classical world line. The difficulties come
from two aspects. One is the mass-shell constraint and the other comes
from the heterogeneous-manifolds that particles and fields reside
on. This study develops a general geometric (manifestly covariant)
field theory for classical relativistic particle-field systems. In
considering the mass-shell constraint, the Euler-Lagrange-Barut (ELB)
equation as a geometric version of the Euler-Lagrange (EL) equation
is applied to determine the world lines of the relativistic particles.
As a differential equation in the standard field theory, the infinitesimal
criterion of the symmetry condition is converted into an integro-differential
equation. To overcome the second difficulty, we develop a weak ELB
equation on the 4D space-time. The weak version of the ELB equation
will play an essential role in establishing the connections between
symmetries and local conservation laws. Using field theory together
with the weak ELB equation developed here, the conservation laws can
be systematically derived from the symmetries that the systems admit.
\end{abstract}
\maketitle

\section{introduction}

Classical relativistic particle-field systems, where many particles
interact with the self-generated and high background fields, are often
encountered in astrophysics \citep{Ma1995,Wei2021,Beklemishev2004},
accelerator physics \citep{Xie2007} and gyrokinetic systems in plasma
physics \citep{Littlejohn1984,Beklemishev2004,Beklemishev2012}. For
these systems, one of the significant topics is focused on the derivation
of energy and momentum conservation laws \citep{Boghosian2003b,Similon1985,Brizard2011a}.

As a widely accepted fundamental principle, conservation laws can
be derive by the corresponding symmetry that the Lagrangian (or action)
of the system admit. This is the so-called Noether's theorem \citep{Noether1918}.
This method has been widely used in deriving the energy-momentum conservation
in quantum field systems by the space-time translation symmetry \citep{Peskin2018}.
However, for classical particle-field systems, the derivation process
is still elusive. For example, for electromagnetic system coupled
with relativistic particles, the energy-momentum conservation was
first derived by Landau and Lifshitz \citep{Landau1971}, and just
reformulated into a geometric form recently from the space-time translation
symmetry \citep{Fan2018}.

Different from the quantum systems, the dynamics of the particles
and the fields reside on heterogeneous manifolds. The fields, e.g.,
the electromagnetic fields, are defined on the 4D space-time domain,
while the particles as world lines in Minkowski space are only defined
on 1D parametric space (e.g, time-axis), which make the standard Euler-Lagrange
(EL) breakdown \citep{Qin2014b}. This difficulty has been overcame
recently by transformed the EL equation into a weak form \citep{Qin2014b,Fan2019b,Fan2020},
and the corresponding method has been applied in non-relativistic
systems, such as the Vlasov-Possion (VP) system, Vlasov-Darwin (VD)
system and gyrokinetic system in plasma physics.

However, for relativistic situation with the geometric setting, this
theory was only reformulated for the Maxwell's system \citep{Fan2018}.
In promoting the method to geometric formalism, the proper-time parameter
rather than time parameter is often used, which yields the mass-shell
constraint. In considering this constraint, the equation of motion,
derived by the Hamilton's principle, will be the Euler-Lagrange-Barut
(ELB) equation instead of the standard EL equation \citep{Barut1964,Brehme1971}.
Besides, the using of proper-time parameter also makes the Lagrangian
density a functional instead a function of particle's world line.
Due to the existing difficulties, for general classical relativistic
particle-field systems, the weak form equation of particle's motion
and infinitesimal criterion of symmetry condition for is still unclear.

In this study, we will extended the theory constructed in Ref.\,\citep{Fan2018}
to general high-order field systems. High-order electromagnetic field
theories always appear in the study of gyrokinetic systems for magnetized
plasma \citep{Qin2007a} and the Podolsky system \citep{Podolsky1942,Bopp1940}
for the radiation reaction of classical charged particles. The weak
ELB equation for Maxwell's system is also extended to general situations.
The infinitesimal criterion of symmetry condition is also reformulated
in considering the geometric setting. Moreover, because the Lagrangian
density is a functional, the criterion we derived here is then a integro-differential
equation. This is quite different from the standard field theory,
in which the Lagrangian density is a function, and the infinitesimal
criterion is consequently a differential equation \citep{Olver1988}.
Using the the general weak ELB equation and the infinitesimal criterion
developed here, the conservation laws can be systematically derived
from the underlying symmetries. As a simple but nontrivial high-order
case, we will calculate the energy-moment conservation laws for Podolsky
system coupled with relativistic charged particles.

The rest of this paper is organized as follows. In Sec.\,\ref{sec:Hamilton's-principle},
we will introduce the action of a general classical relativistic particle-field
systems and the weak ELB equation is established as well. In Sec.\,\ref{sec:infinitesimal-criterion},
the geometric infinitesimal criterion of the symmetry is developed
as required by the mass-shell constraint. Using the geometric weak
Euler-Lagrange equation and geometric infinitesimal criterion, the
conservation laws are obtained in Sec.\,\ref{sec:conservation-laws}.
In the last section, we will derive the energy-momentum conservation
laws for high-order electromagnetic systems coupled with charged particles.

\section{classical relativistic particle-field systems and geometric weak
Euler-Lagrange-Barut equation \label{sec:Hamilton's-principle}}

We start from the geometric action of the particle-field system and
extend the theory developed in Ref.\,\citep{Fan2018,Fan2019b} to
the general relativistic situation. The geometric action of a classical
relativistic particle-field system is generally written as

\begin{equation}
\mathcal{A}=\sum_{a}\int_{a_{1}}^{a_{2}}L_{a}\left(\tau_{a},\boldsymbol{\chi}_{a}\left(\tau_{a}\right),\dot{\boldsymbol{\chi}}_{a}\left(\tau_{a}\right),\mathrm{pr}^{(n)}\boldsymbol{\psi}\left(\boldsymbol{\chi}_{a}\left(\tau_{a}\right)\right)\right)d\tau_{a}+\int_{\Omega}\mathcal{L}_{F}\left(\boldsymbol{\chi},\mathrm{pr}^{(n)}\boldsymbol{\psi}\left(\boldsymbol{\chi}\right)\right)d^{4}\boldsymbol{\chi},\label{eq:action-1}
\end{equation}
where the subscript $a$ labels particles, $\boldsymbol{\chi}$ is
the space-time position, $\boldsymbol{\psi}\left(\boldsymbol{\chi}\right)$
is a vector (or any tensor) field defined on space-time. Here, $\mathrm{pr}^{(n)}\boldsymbol{\psi}\left(\boldsymbol{\chi}\right)$
is the prolongation of the field $\boldsymbol{\psi}\left(\boldsymbol{\chi}\right)$,
which contains $\boldsymbol{\psi}\left(\boldsymbol{\chi}\right)$
and its derivatives up to the $n$th order, i.e.,
\begin{equation}
\mathrm{pr}^{(n)}\boldsymbol{\psi}\left(\boldsymbol{\chi}\right)=\left(\boldsymbol{\psi}\left(\boldsymbol{\chi}\right),\partial_{\mu_{1}}\boldsymbol{\psi}\left(\boldsymbol{\chi}\right),\cdots,\partial_{\mu_{1}}\partial_{\mu_{2}}\cdots\partial_{\mu_{n}}\boldsymbol{\psi}\left(\boldsymbol{\chi}\right)\right),\label{eq:prolongation-=00005Cpsi}
\end{equation}
where $\partial_{\mu_{i}}$ $\left(i=1,2,\cdots,n\right)$ represents
derivative with respect to space time coordinates. In Eq.\,(\ref{eq:action-1}),
$\boldsymbol{\chi}_{a}\left(\tau_{a}\right)$ is the world line of
the $a$th particle and $\tau_{a}$ is the proper time parameter.
$\dot{\boldsymbol{\chi}}_{a}\left(\tau_{a}\right)\equiv d\boldsymbol{\chi}_{a}\left(\tau_{a}\right)/d\tau_{a}$
is the 4-velocity of the $a$th particle which satisfy the mass-shell
constraint 
\begin{equation}
\dot{\chi}_{a}^{\mu}\dot{\chi}_{a\mu}=c^{2},\label{eq:mass-shell}
\end{equation}
where the Lorentzian metric $\eta_{\mu\nu}=\mathrm{diag}\left\{ 1,-1,-1,-1\right\} $
is applied to define $\dot{\chi}_{a\mu}$, i.e., $\dot{\chi}_{a\mu}=\eta_{\mu\nu}\dot{\chi}_{a}^{\nu}$.
Due to the subsidiary condition (\ref{eq:mass-shell}), the variation
of the action $\mathcal{A}$ with respect to $\delta\boldsymbol{\chi}_{a}$
won't yields the standard Euler-Lagrange equation but the equation
as follows
\begin{equation}
E_{\boldsymbol{\chi}_{a}\mu}\left(L_{a}\right)=0,\label{eq:ELB}
\end{equation}
where the linear operator $E_{\boldsymbol{\chi}_{a}\mu}$ is defined
by
\begin{equation}
E_{\boldsymbol{\chi}_{a}\mu}\coloneqq\frac{\partial}{\partial\chi_{a}^{\mu}}-\frac{d}{d\tau_{a}}\left[\frac{\partial}{\partial\dot{\chi}_{a}^{\mu}}+\frac{1}{c^{2}}\left(\mathrm{id}-\dot{\chi}_{a}^{\nu}\frac{\partial}{\partial\dot{\chi}_{a}^{\nu}}\right)\right],\label{eq:Euler-operator-=00005Cchi}
\end{equation}
which is different from the standard Euler operator by the existence
of the last two terms. To be faithful to history, equation $\left(\ref{eq:ELB}\right)$
was first obtained by Barut using the Lagrange multiplier method \citep{Barut1964}.
Therefore, we will refer to the linear operator $E_{\boldsymbol{\chi}_{a}\mu}$
as the Euler-Lagrange-Barut (ELB) operator and call equation (\ref{eq:ELB})
the ELB equation.

The integral of the $a$th particle's Lagrangian $L_{a}$ in Eq.\,(\ref{eq:action-1})
is along an arbitrary time-like world line (denoted by $l_{a}$) which
connected two fixed world points $a_{1}$ and $a_{2}$ at the space-time
$\mathbb{R}^{4}$, while the integral of the field's Lagrangian density
$\mathcal{L}_{F}$ is over the the space-time domain $\Omega$. Hence,
the integral of the Lagrangian density $\mathcal{L}_{F}$ for the
field $\boldsymbol{\psi}$ is over space-time, and the integral of
Lagrangian $L_{a}$ for the $a$th particle is over proper time only.
As a consequence, the action in the form (\ref{eq:action-1}) is not
easily applicable for Noether\textquoteright s procedure of deriving
local conservation laws.

To deal with this problem, we multiply the first part on the right-hand
side of Eq.\,(\ref{eq:action-1}) by the following identity
\begin{equation}
\int\delta_{a}d^{4}\boldsymbol{\chi}=1,\label{eq:integral-delta}
\end{equation}
where $\delta_{a}\equiv\delta\left(\boldsymbol{\chi}-\boldsymbol{\chi}_{a}\left(\tau_{a}\right)\right)$
is the Dirac's delta function. The action $\mathcal{A}$ in Eq.\,(\ref{eq:action-1})
is then transformed into one integral over space-time,
\begin{align}
 & \mathcal{A}=\int_{\Omega}\mathcal{L}\left(\boldsymbol{\chi},\left[\boldsymbol{\chi}_{a}\right],\mathrm{pr}^{(n)}\boldsymbol{\psi}\left(\boldsymbol{\chi}\right)\right)d^{4}\boldsymbol{\chi},\label{eq:action-2}
\end{align}
where 
\begin{align}
 & \mathcal{L}=\sum_{a}\mathcal{L}_{a}+\mathcal{L}_{F}\left(\boldsymbol{\chi},\mathrm{pr}^{(n)}\boldsymbol{\psi}\left(\boldsymbol{\chi}\right)\right),\label{eq:Lagrangian-density}\\
 & \mathcal{L}_{a}=\int_{a_{1}}^{a_{2}}\ell_{a}d\tau_{a},\;\ell_{a}=\ell_{a}\left(\tau_{a},\boldsymbol{\chi},\mathrm{pr}^{(1)}\boldsymbol{\chi}_{a}\left(\tau_{a}\right),\mathrm{pr}^{(n)}\boldsymbol{\psi}\left(\boldsymbol{\chi}\right)\right)=L_{a}\delta_{a}.\label{eq:Lagrangian-density-Particle}
\end{align}
Different from the non-geometric situation, the Lagrangian density
$\mathcal{L}$ is not a function but a functional of the particle's
world line. To differ from the other local variables (such as $\boldsymbol{\chi}$
and $\boldsymbol{\psi}\left(\boldsymbol{\chi}\right)$), we enclose
$\boldsymbol{\chi}_{a}$ by square brackets.

We now calculate how the action (\ref{eq:action-2}) varies in response
to the variations of $\delta\boldsymbol{\chi}_{a}$ and $\delta\boldsymbol{\psi}$,
\begin{equation}
\delta\mathcal{A}=\sum_{a}\int_{a_{1}}^{a_{2}}\left[\int_{\Omega}E_{\boldsymbol{\chi}_{a}\mu}\left(\ell_{a}\right)d^{4}\boldsymbol{\chi}\right]\delta\chi_{a}^{\mu}d\tau_{a}+\int_{\Omega}E_{\boldsymbol{\psi}}\left(\mathcal{L}\right)\cdot\delta\boldsymbol{\psi}d^{4}\boldsymbol{\chi},\label{eq:delta-A}
\end{equation}
where
\begin{equation}
E_{\boldsymbol{\psi}}=\frac{\partial}{\partial\boldsymbol{\psi}}+\sum_{j=1}^{n}\left(-1\right)^{j}D_{\mu_{1}}\cdots D_{\mu_{j}}\frac{\partial}{\partial\left(\partial_{\mu_{1}}\cdots\partial_{\mu_{j}}\boldsymbol{\psi}\right)}\label{eq:Euler-operator-=00005Cpsi}
\end{equation}
is the Euler operator for the field $\boldsymbol{\psi}$. Here, $\mathrm{id}$
in Eq.\,(\ref{eq:Euler-operator-=00005Cchi}) is the identity operator.
Using Hamilton's principle, we immediately obtain the equation of
motion for particles and fields
\begin{align}
 & \int_{\Omega}E_{\boldsymbol{\chi}_{a}\mu}\left(\ell_{a}\right)d^{4}\boldsymbol{\chi}=0,\label{eq:submanifold-EL-equation}\\
 & E_{\boldsymbol{\psi}}\left(\mathcal{L}\right)=0.\label{eq:Psi-EL-equation}
\end{align}
Similar to the submanifold EL equation given in Refs.\,\citep{Qin2014b,Fan2019b},
equation (\ref{eq:submanifold-EL-equation}) is called the submanifold
ELB equation. Using the linear property of ELB operator, we can easily
prove that the submanifold ELB equation (\ref{eq:submanifold-EL-equation})
is equivalent to ELB equation (\ref{eq:ELB}).

To apply equation (\ref{eq:submanifold-EL-equation}) to Noether's
method, we need the explicit expression of $E_{\boldsymbol{\chi}_{a}\mu}\left(\ell_{a}\right)$.
For electromagnetic system, the expression was given in the previous
work \citep{Fan2018}. We now derive a general expression of $E_{\boldsymbol{\chi}_{a}\mu}\left(\ell_{a}\right)$.
We first transform the first and the last three terms of Eq.\,(\ref{eq:submanifold-EL-equation})
into following forms
\begin{align}
 & \frac{\partial\ell_{a}}{\partial\chi_{a}^{\mu}}=\frac{\partial}{\partial\chi_{a}^{\mu}}\left(L_{a}\delta_{a}\right)=\frac{\partial\delta_{a}}{\partial\chi_{a}^{\mu}}L_{a}+\frac{\partial L_{a}}{\partial\chi_{a}^{\mu}}\delta_{a}=-\frac{\partial\delta_{a}}{\partial\chi^{\mu}}L_{a}+\frac{\partial L_{a}}{\partial\chi_{a}^{\mu}}\delta_{a}\nonumber \\
 & =-\frac{D}{D\chi^{\mu}}\left(L_{a}\delta_{a}\right)+\frac{\partial L_{a}}{\partial\chi_{a}^{\mu}}\delta_{a}=\frac{D}{D\chi^{\nu}}\left(-L_{a}\delta_{a}\eta_{\:\mu}^{\nu}\right)+\frac{\partial L_{a}}{\partial\chi_{a}^{\mu}}\delta_{a}.\label{eq:first-term}
\end{align}
\begin{align}
 & -\frac{d}{d\tau_{a}}\left\{ \left[\frac{\partial L_{a}}{\partial\dot{\chi}_{a}^{\mu}}+\frac{1}{c^{2}}\dot{\chi}_{a\mu}\left(L_{a}-\dot{\chi}_{a}^{\sigma}\frac{\partial L_{a}}{\partial\dot{\chi}_{a}^{\sigma}}\right)\right]\delta_{a}\right\} \nonumber \\
 & =-\frac{d\delta_{a}}{d\tau_{a}}\left[\frac{\partial L_{a}}{\partial\dot{\chi}_{a}^{\mu}}+\frac{1}{c^{2}}\dot{\chi}_{a\mu}\left(L_{a}-\dot{\chi}_{a}^{\sigma}\frac{\partial L_{a}}{\partial\dot{\chi}_{a}^{\sigma}}\right)\right]-\frac{d}{d\tau_{a}}\left[\frac{\partial L_{a}}{\partial\dot{\chi}_{a}^{\mu}}+\frac{1}{c^{2}}\dot{\chi}_{a\mu}\left(L_{a}-\dot{\chi}_{a}^{\sigma}\frac{\partial L_{a}}{\partial\dot{\chi}_{a}^{\sigma}}\right)\right]\delta_{a}.\label{eq:second-term}
\end{align}
The first term on the right-hand side of Eq.\,(\ref{eq:second-term})
can be rewritten as
\begin{align}
 & -\frac{d\delta_{a}}{d\tau_{a}}\left[\frac{\partial L_{a}}{\partial\dot{\chi}_{a}^{\mu}}+\frac{1}{c^{2}}\dot{\chi}_{a\mu}\left(L_{a}-\dot{\chi}_{a}^{\sigma}\frac{\partial L_{a}}{\partial\dot{\chi}_{a}^{\sigma}}\right)\right]=-\dot{\chi}_{a}^{\nu}\frac{\partial\delta_{a}}{\partial\chi_{a}^{\nu}}\left[\frac{\partial L_{a}}{\partial\dot{\chi}_{a}^{\mu}}+\frac{1}{c^{2}}\dot{\chi}_{a\mu}\left(L_{a}-\dot{\chi}_{a}^{\sigma}\frac{\partial L_{a}}{\partial\dot{\chi}_{a}^{\sigma}}\right)\right]\nonumber \\
 & =\frac{D}{D\chi^{\nu}}\left\{ \dot{\chi}_{a}^{\nu}\left[\frac{\partial\ell_{a}}{\partial\dot{\chi}_{a}^{\mu}}+\frac{1}{c^{2}}\dot{\chi}_{a\mu}\left(\ell_{a}-\dot{\chi}_{a}^{\sigma}\frac{\partial\ell_{a}}{\partial\dot{\chi}_{a}^{\sigma}}\right)\right]\right\} .\label{eq:16}
\end{align}
Substituting Eqs.\,(\ref{eq:first-term})-(\ref{eq:16}) into $E_{\boldsymbol{\chi}_{a}\mu}\left(\ell_{a}\right)$,
we have 
\begin{equation}
E_{\boldsymbol{\chi}_{a}\mu}\left(\ell_{a}\right)=\frac{D}{D\chi^{\nu}}\left\{ -\ell_{a}\eta_{\:\mu}^{\nu}+\dot{\chi}_{a}^{\nu}\left[\frac{\partial\ell_{a}}{\partial\dot{\chi}_{a}^{\mu}}+\frac{1}{c^{2}}\dot{\chi}_{a\mu}\left(\ell_{a}-\dot{\chi}_{a}^{\sigma}\frac{\partial\ell_{a}}{\partial\dot{\chi}_{a}^{\sigma}}\right)\right]\right\} ,\label{eq:weak-ELB}
\end{equation}
where we used the ELB equation (\ref{eq:ELB}). We will refer to Eq.\,(\ref{eq:weak-ELB})
as the weak ELB equation, which as a differential equation is equivalent
to the submanifold ELB equation (\ref{eq:submanifold-EL-equation}).
Just like the weak EL equation used in non-relativistic particle-field
systems, the weak ELB equation is also elemental in deriving local
conservation laws for relativistic situation.

\section{geometric infinitesimal criterion of symmetry condition \label{sec:infinitesimal-criterion}}

We now turn to discuss the symmetries of the relativistic particle-field
systems. A symmetry of the action $\mathcal{A}$ is a group of transformation
\begin{align}
 & g_{\epsilon}\cdot\left(\tau_{a},\boldsymbol{\chi},\boldsymbol{\chi}_{a}\left(\tau_{a}\right),\boldsymbol{\psi}\left(\boldsymbol{\chi}\right)\right)\equiv\left(Z_{a\epsilon}\left(\tau_{a},\boldsymbol{\chi},\boldsymbol{\chi}_{a},\boldsymbol{\psi}\right),\boldsymbol{\Xi}_{\epsilon}\left(\tau_{a},\boldsymbol{\chi},\boldsymbol{\chi}_{a},\boldsymbol{\psi}\right),\right.\nonumber \\
 & \left.\vphantom{\boldsymbol{\Phi}_{\epsilon}}\boldsymbol{\Theta}_{a\epsilon}\left(\tau_{a},\boldsymbol{\chi},\boldsymbol{\chi}_{a},\boldsymbol{\psi}\right),\boldsymbol{\Phi}_{\epsilon}\left(\tau_{a},\boldsymbol{\chi},\boldsymbol{\chi}_{a},\boldsymbol{\psi}\right)\right)=\left(\tilde{\tau}_{a},\tilde{\boldsymbol{\chi}},\tilde{\boldsymbol{\chi}}_{a},\tilde{\boldsymbol{\psi}}\left(\tilde{\boldsymbol{\chi}}\right)\right)\label{eq:group-transformation}
\end{align}
such that
\begin{align}
 & \int_{\Omega}\mathcal{L}\left(\boldsymbol{\chi},\left[\boldsymbol{\chi}_{a}\right],\mathrm{pr}^{(n)}\boldsymbol{\psi}\left(\boldsymbol{\chi}\right)\right)d^{4}\boldsymbol{\chi}=\int_{\tilde{\Omega}}\mathcal{L}\left(\tilde{\tau}_{a},\tilde{\boldsymbol{\chi}},\left[\tilde{\boldsymbol{\chi}}_{a}\right],\tilde{\boldsymbol{\psi}}\left(\tilde{\boldsymbol{\chi}}\right)\right)d^{4}\tilde{\boldsymbol{\chi}},\label{eq:symmetry-condition}
\end{align}
and
\begin{equation}
\dot{\chi}_{a}^{\mu}\dot{\chi}_{a\mu}=\dot{\tilde{\chi}}_{a}^{\mu}\dot{\tilde{\chi}}_{a\mu}=c^{2},\label{eq:symmetry-constraint}
\end{equation}
where $g_{\epsilon}=\left(Z_{a\epsilon},\boldsymbol{\Xi}_{\epsilon},\boldsymbol{\Theta}_{a\epsilon},\boldsymbol{\Phi}_{\epsilon}\right)$
constitutes a continuous group of transformations parameterized by
$\epsilon$. Equations (\ref{eq:symmetry-condition}) and (\ref{eq:symmetry-constraint})
are called the symmetry conditions. Different from the standard field
systems, the second condition (\ref{eq:symmetry-constraint}) is needed
now due to the mass-shell constraint.

To derive the local conservation laws, an infinitesimal version of
the symmetry condition is required. For this purpose, we take the
derivative of Eqs.\,(\ref{eq:symmetry-condition}) and (\ref{eq:symmetry-constraint})
with respect to $\epsilon$ at $\epsilon=0$,
\begin{equation}
\frac{d}{d\epsilon}\bigg|_{0}\int_{\tilde{\Omega}}\left[\sum_{a}\int_{\tilde{a}_{1}}^{\tilde{a}_{2}}\ell_{a}\left(\tilde{\tau}_{a},\tilde{\boldsymbol{\chi}},\mathrm{pr}^{(1)}\tilde{\boldsymbol{\chi}}_{a}\left(\tilde{\tau}_{a}\right),\mathrm{pr}^{(n)}\tilde{\boldsymbol{\psi}}\left(\tilde{\boldsymbol{\chi}}\right)\right)d\tilde{\tau}_{a}+\mathcal{L}_{F}\left(\tilde{\boldsymbol{\chi}},\mathrm{pr}^{(n)}\tilde{\boldsymbol{\psi}}\left(\tilde{\boldsymbol{\chi}}\right)\right)\right]d^{4}\tilde{\boldsymbol{\chi}}=0.\label{eq:d/d=00005Cepsilon}
\end{equation}
\begin{equation}
\frac{d}{d\epsilon}\bigg|_{0}\left[\dot{\tilde{\chi}}_{a}^{\mu}\dot{\tilde{\chi}}_{a\mu}\right]=0\label{eq:mass-shell-epsilon}
\end{equation}
Substituting Eqs.\,(\ref{eq:Lagrangian-density}) and (\ref{eq:Lagrangian-density-Particle})
into Eq.\,(\ref{eq:d/d=00005Cepsilon}), we have
\begin{align}
 & \int_{\Omega}\frac{d}{d\epsilon}\bigg|_{0}\left\{ \left[\sum_{a}\int_{a_{1}}^{a_{2}}\ell_{a}\left(\tilde{\tau}_{a},\tilde{\boldsymbol{\chi}},\mathrm{pr}^{(1)}\tilde{\boldsymbol{\chi}}_{a}\left(\tilde{\tau}_{a}\right),\mathrm{pr}^{(n)}\tilde{\boldsymbol{\psi}}\left(\tilde{\boldsymbol{\chi}}\right)\right)j_{a}\left(\epsilon\right)d\tau_{a}\right.\right.\nonumber \\
 & \left.\vphantom{\int_{\tilde{a}_{1}}^{\tilde{a}_{2}}}\left.\vphantom{\int_{\tilde{a}_{1}}^{\tilde{a}_{2}}}+\mathcal{L}_{F}\left(\tilde{\boldsymbol{\chi}},\mathrm{pr}^{(n)}\tilde{\boldsymbol{\psi}}\left(\tilde{\boldsymbol{\chi}}\right)\right)\right]\mathrm{det}\boldsymbol{J}\left(\epsilon\right)\right\} d^{4}\boldsymbol{\chi}=0,\label{eq:22}
\end{align}
where 
\begin{equation}
j_{a}\left(\epsilon\right)\equiv\frac{d\tilde{\tau}_{a}}{d\tau_{a}},\;\boldsymbol{J}\left(\epsilon\right)\equiv\frac{\partial\tilde{\boldsymbol{\chi}}}{\partial\boldsymbol{\chi}},\label{eq:j_a-J}
\end{equation}
Because equation (\ref{eq:22}) survive for any small integral domains,
the integrand must be zero, i.e., 
\begin{align}
 & \frac{d}{d\epsilon}\bigg|_{0}\left[\sum_{a}\int_{a_{1}}^{a_{2}}\ell_{a}\left(\tilde{\tau}_{a},\tilde{\boldsymbol{\chi}},\mathrm{pr}^{(1)}\tilde{\boldsymbol{\chi}}_{a}\left(\tilde{\tau}_{a}\right),\mathrm{pr}^{(n)}\tilde{\boldsymbol{\psi}}\left(\tilde{\boldsymbol{\chi}}\right)\right)j_{a}\left(\epsilon\right)d\tau_{a}\right.\nonumber \\
 & \left.\vphantom{\int_{\tilde{a}_{1}}^{\tilde{a}_{2}}}+\mathcal{L}_{F}\left(\tilde{\boldsymbol{\chi}},\mathrm{pr}^{(n)}\tilde{\boldsymbol{\psi}}\left(\tilde{\boldsymbol{\chi}}\right)\right)\right]\mathrm{det}\boldsymbol{J}\left(\epsilon\right)=0.\label{eq:25}
\end{align}
Equation (\ref{eq:25}) can be finally transformed into
\begin{equation}
\sum_{a}\int_{a_{1}}^{a_{2}}\left[\mathrm{pr}^{\left(n\right)}\boldsymbol{v}\left(\ell_{a}\right)+\ell_{a}\dot{\zeta}_{a}\right]d\tau_{a}+\mathrm{pr}^{\left(n\right)}\boldsymbol{v}\left(\mathcal{L}_{F}\right)+\mathcal{L}D_{\mu}\xi^{\mu}=0,\label{eq:infinitesimal-criterion-1}
\end{equation}
where
\begin{equation}
\boldsymbol{v}=\frac{d}{d\epsilon}\bigg|_{0}g_{\epsilon}\cdot\left(\tau_{a},\boldsymbol{\chi},\boldsymbol{\chi}_{a},\boldsymbol{\psi}\right)=\sum_{a}\zeta_{a}\frac{\partial}{\partial\tau_{a}}+\boldsymbol{\xi}\cdot\frac{\partial}{\partial\boldsymbol{\chi}}+\sum_{a}\boldsymbol{\theta}_{a}\cdot\frac{\partial}{\partial\boldsymbol{\chi}_{a}}+\boldsymbol{\phi}\cdot\frac{\partial}{\partial\boldsymbol{\psi}}\label{eq:infinitesimal-generator}
\end{equation}
is the infinitesimal generator of the group transformation, and where
\begin{equation}
\begin{cases}
\zeta_{a}=\zeta_{a}\left(\tau_{a},\boldsymbol{\chi},\boldsymbol{\chi}_{a},\boldsymbol{\psi}\right)=\frac{d}{d\epsilon}\bigg|_{0}Z_{a\epsilon}\left(\tau_{a},\boldsymbol{\chi},\boldsymbol{\chi}_{a},\boldsymbol{\psi}\right),\\
\boldsymbol{\xi}=\boldsymbol{\xi}\left(\tau_{a},\boldsymbol{\chi},\boldsymbol{\chi}_{a},\boldsymbol{\psi}\right)=\frac{d}{d\epsilon}\bigg|_{0}\boldsymbol{\Xi}_{\epsilon}\left(\tau_{a},\boldsymbol{\chi},\boldsymbol{\chi}_{a},\boldsymbol{\psi}\right),\\
\boldsymbol{\theta}_{a}=\boldsymbol{\theta}_{a}\left(\tau_{a},\boldsymbol{\chi},\boldsymbol{\chi}_{a},\boldsymbol{\psi}\right)=\frac{d}{d\epsilon}\bigg|_{0}\boldsymbol{\Theta}_{a\epsilon}\left(\tau_{a},\boldsymbol{\chi},\boldsymbol{\chi}_{a},\boldsymbol{\psi}\right),\\
\boldsymbol{\phi}=\boldsymbol{\phi}\left(\tau_{a},\boldsymbol{\chi},\boldsymbol{\chi}_{a},\boldsymbol{\psi}\right)=\frac{d}{d\epsilon}\bigg|_{0}\boldsymbol{\Phi}_{\epsilon}\left(\tau_{a},\boldsymbol{\chi},\boldsymbol{\chi}_{a},\boldsymbol{\psi}\right).
\end{cases}
\end{equation}
Here, $\mathrm{pr}^{\left(n\right)}\boldsymbol{v}$ is the prolongation
of $\boldsymbol{v}$ defined by 
\begin{align}
 & \mathrm{pr}^{\left(n\right)}\boldsymbol{v}=\frac{d}{d\epsilon}\bigg|_{0}\left(\tilde{\tau}_{a},\tilde{\boldsymbol{\chi}},\mathrm{pr}^{\left(1\right)}\tilde{\boldsymbol{\chi}}_{a},\mathrm{pr}^{\left(n\right)}\tilde{\boldsymbol{\psi}}\left(\tilde{\boldsymbol{\chi}}\right)\right)\nonumber \\
 & =\boldsymbol{v}+\sum_{a}\theta_{a1}^{\mu}\frac{\partial}{\partial\dot{\chi}_{a}^{\mu}}+\sum_{k=1}^{n}\phi_{\mu_{1}\dots\mu_{k}}^{\alpha}\frac{\partial}{\partial\left(D_{\mu_{1}}\cdots D_{\mu_{k}}\psi^{\alpha}\right)},\label{eq:prolongation-defination}
\end{align}
where 
\begin{align}
 & \begin{cases}
\theta_{a1}^{\mu}=\zeta_{a}\ddot{\chi}_{a}^{\mu}+\dot{q}_{a}^{\mu},\\
\phi_{\mu_{1}\dots\mu_{k}}^{\alpha}=\xi^{\mu}D_{\mu_{1}}\cdots D_{\mu_{k}}\left(D_{\mu}\psi^{\alpha}\right)+D_{\mu_{1}}\cdots D_{\mu_{k}}Q^{\alpha},
\end{cases}
\end{align}
and 
\begin{equation}
\begin{cases}
\boldsymbol{q}_{a}\equiv\boldsymbol{\theta}_{a}-\zeta_{a}\dot{\boldsymbol{\chi}}_{a},\\
Q^{\alpha}\equiv\phi^{\alpha}-\xi^{\mu}D_{\mu}\psi^{\alpha},
\end{cases}\label{eq:characteristic}
\end{equation}
are characteristics of the infinitesimal generator $\boldsymbol{v}$.
Equation (\ref{eq:infinitesimal-criterion-1}) is the infinitesimal
criterion of the symmetry condition (\ref{eq:symmetry-condition}).
Detailed derivation process of the prolongation formula (\ref{eq:prolongation-defination})
can be found in Ref.\,\citep{Olver1988}. Different from the situation
in the previous references, an integral along the particle's world
line appeared in the infinitesimal criterion (see Eq.\,(\ref{eq:infinitesimal-criterion-1})).
This originally comes from the fact that the Lagrangian density $\mathcal{L}$
is a functional rather than a function of particle's world lines.
Owing to this integral, the infinitesimal criterion (\ref{sec:infinitesimal-criterion})
is not a differential equation but a integro-differential equation.

Similarly, using Eq.\,(\ref{eq:mass-shell-epsilon}), we can obtained
another infinitesimal criterion as
\begin{equation}
\dot{q}_{a}^{\mu}\dot{\chi}_{a\mu}=0.\label{eq:infinitesimal-criterion-2}
\end{equation}
In deriving Eq.\,(\ref{eq:infinitesimal-criterion-2}), we used the
following equations 
\begin{align}
 & \frac{d}{d\epsilon}\bigg|_{0}\dot{\tilde{\chi}}^{\mu}=\theta_{a1}^{\mu}=\zeta_{a}\ddot{\chi}_{a}^{\mu}+\dot{q}_{a}^{\mu},\\
 & \ddot{\chi}_{a}^{\mu}\dot{\chi}_{a\mu}\equiv0.
\end{align}

\section{conservation laws \label{sec:conservation-laws}}

Due to the particularities of the infinitesimal criterion shown in
Sec.\,\ref{sec:infinitesimal-criterion}, especially the existence
of the additional infinitesimal criterion (\ref{eq:infinitesimal-criterion-2}),
the derivation process of local conservation law and the and the final
results are quite different from the standard situation without constraints.
We next show how the infinitesimal criterion (\ref{eq:infinitesimal-criterion-1})
and (\ref{eq:infinitesimal-criterion-2}) determine a conservation
law.

We first transform Eq.\,(\ref{eq:infinitesimal-criterion-1}) into
another equivalent form. The first term in Eq.\,(\ref{eq:infinitesimal-criterion-1})
can be rewritten by components as 
\begin{align}
 & \mathrm{pr}^{\left(n\right)}\boldsymbol{v}\left(\ell_{a}\right)+\ell_{a}\dot{\zeta}_{a}=\zeta_{a}\frac{\partial\ell_{a}}{\partial\tau_{a}}+\theta_{a}^{\mu}\frac{\partial\ell_{a}}{\partial\chi_{a}^{\mu}}+\theta_{a1}^{\mu}\frac{\partial\ell_{a}}{\partial\dot{\chi}_{a}^{\mu}}+\ell_{a}\dot{\zeta}_{a}\nonumber \\
 & +\xi^{\mu}\frac{\partial\ell_{a}}{\partial\chi^{\mu}}+\phi^{\alpha}\frac{\partial\ell_{a}}{\partial\psi^{\alpha}}+\sum_{k=1}^{n}\phi_{\mu_{1}\dots\mu_{k}}^{\alpha}\frac{\partial\ell_{a}}{\partial\left(D_{\mu_{1}}\cdots D_{\mu_{k}}\psi^{\alpha}\right)}.\label{eq:pr-v-l_a}
\end{align}
From the first four terms of right-hand side of Eq.\,(\ref{eq:pr-v-l_a}),
we have
\begin{align}
 & \zeta_{a}\frac{\partial\ell_{a}}{\partial\tau_{a}}+\theta_{a}^{\mu}\frac{\partial\ell_{a}}{\partial\chi_{a}^{\mu}}+\left(\zeta_{a}\ddot{\chi}_{a}^{\mu}+\dot{q}_{a}^{\mu}\right)\frac{\partial\ell_{a}}{\partial\dot{\chi}_{a}^{\mu}}+\ell_{a}\dot{\zeta}_{a}\nonumber \\
 & =\zeta_{a}\left(\frac{\partial\ell_{a}}{\partial\tau_{a}}+\dot{\chi}_{a}^{\mu}\frac{\partial\ell_{a}}{\partial\chi_{a}^{\mu}}+\ddot{\chi}_{a}^{\mu}\frac{\partial\ell_{a}}{\partial\dot{\chi}_{a}^{\mu}}\right)+q_{a}^{\mu}\frac{\partial\ell_{a}}{\partial\chi_{a}^{\mu}}+\dot{q}_{a}^{\mu}\frac{\partial\ell_{a}}{\partial\dot{\chi}_{a}^{\mu}}+\ell_{a}\dot{\zeta}_{a}\nonumber \\
 & =\zeta_{a}\frac{d\ell_{a}}{d\tau_{a}}+\ell_{a}\dot{\zeta}_{a}+\frac{d}{d\tau_{a}}\left(q_{a}^{\mu}\frac{\partial\ell_{a}}{\partial\dot{\chi}_{a}^{\mu}}\right)+q_{a}^{\mu}\left(\frac{\partial\ell_{a}}{\partial\chi_{a}^{\mu}}-\frac{d}{d\tau_{a}}\frac{\partial\ell_{a}}{\partial\dot{\chi}_{a}^{\mu}}\right)\nonumber \\
 & =\frac{d}{d\tau_{a}}\left(\ell_{a}\zeta_{a}\right)+\frac{d}{d\tau_{a}}\left[q_{a}^{\mu}\frac{\partial\ell_{a}}{\partial\dot{\chi}_{a}^{\mu}}\right]+\frac{q_{a}^{\mu}}{c^{2}}\frac{d}{d\tau_{a}}\left[\left(\ell_{a}-\frac{\partial\ell_{a}}{\partial\dot{\chi}_{a}^{\nu}}\dot{\chi}_{a}^{\nu}\right)\dot{\chi}_{a\mu}\right]+q_{a}^{\mu}\boldsymbol{E}_{\boldsymbol{\chi}_{a}\mu}\left(\ell_{a}\right)\nonumber \\
 & =\frac{d}{d\tau_{a}}\left[\ell_{a}\zeta_{a}+q_{a}^{\mu}\frac{\partial\ell_{a}}{\partial\dot{\chi}_{a}^{\mu}}+\frac{q_{a}^{\mu}}{c^{2}}\left(\ell_{a}-\frac{\partial\ell_{a}}{\partial\dot{\chi}_{a}^{\nu}}\dot{\chi}_{a}^{\nu}\right)\dot{\chi}_{a\mu}\right]-\frac{1}{c^{2}}\left(\ell_{a}-\frac{\partial\ell_{a}}{\partial\dot{\chi}_{a}^{\nu}}\dot{\chi}_{a}^{\nu}\right)\dot{q}_{a}^{\mu}\dot{\chi}_{a\mu}+q_{a}^{\mu}\boldsymbol{E}_{\boldsymbol{\chi}_{a}\mu}\left(\ell_{a}\right)\nonumber \\
 & =\frac{d}{d\tau_{a}}\left[\ell_{a}\zeta_{a}+q_{a}^{\mu}\frac{\partial\ell_{a}}{\partial\dot{\chi}_{a}^{\mu}}+\frac{q_{a}^{\mu}}{c^{2}}\left(\ell_{a}-\frac{\partial\ell_{a}}{\partial\dot{\chi}_{a}^{\nu}}\dot{\chi}_{a}^{\nu}\right)\dot{\chi}_{a\mu}\right]+q_{a}^{\mu}\boldsymbol{E}_{\boldsymbol{\chi}_{a}\mu}\left(\ell_{a}\right),\label{eq:first-three-pr-v-l_a}
\end{align}
where we used the second infinitesimal criterion (\ref{eq:infinitesimal-criterion-2})
in the last step. The last three terms of Eq.\,(\ref{eq:pr-v-l_a})
can be transformed into 
\begin{align}
 & \xi^{\mu}\frac{\partial\ell_{a}}{\partial\chi^{\mu}}+\phi^{\alpha}\frac{\partial\ell_{a}}{\partial\psi^{\alpha}}+\sum_{k=1}^{n}\phi_{\mu_{1}\dots\mu_{k}}^{\alpha}\frac{\partial\ell_{a}}{\partial\left(D_{\mu_{1}}\cdots D_{\mu_{k}}\psi^{\alpha}\right)}\nonumber \\
 & =\xi^{\mu}D_{\mu}\ell_{a}+D_{\mu}\mathbb{P}_{a}^{\mu}+\boldsymbol{E}_{\boldsymbol{\psi}}\left(\ell_{a}\right)\cdot\boldsymbol{Q}\label{eq:last-three}
\end{align}
by the standard derivation process (see\textbf{ }Ref.\,\citep{Olver1988}),
where
\begin{equation}
\mathbb{P}_{a}^{\mu}=\sum_{i=1}^{n}\sum_{j=1}^{i}\left(-1\right)^{j+1}D_{\mu_{j+1}}\cdots D_{\mu_{k}}Q^{\alpha}\left[D_{\mu_{1}}\cdots D_{\mu_{j-1}}\frac{\partial\ell_{a}}{\partial\left(D_{\mu_{1}}\cdots D_{\mu_{j-1}}D_{\mu}D_{\mu_{j+1}}\cdots D_{\mu_{i}}\psi^{\alpha}\right)}\right].\label{eq:P_mu}
\end{equation}
Similarly, the last two terms in Eq.\,(\ref{eq:pr-v-l_a}), (\ref{eq:infinitesimal-criterion-1})
can be written as 
\begin{equation}
\mathrm{pr}^{\left(n\right)}\boldsymbol{v}\left(\mathcal{L}_{F}\right)+\mathcal{L}D_{\mu}\xi^{\mu}=\xi^{\mu}D_{\mu}\mathcal{L}_{F}+\mathcal{L}D_{\mu}\xi^{\mu}+D_{\mu}\mathbb{P}_{F}^{\mu}+\boldsymbol{E}_{\boldsymbol{\psi}}\left(\mathcal{L}_{F}\right)\cdot\boldsymbol{Q},\label{eq:prv-L_F+L}
\end{equation}
where 
\begin{equation}
\mathbb{P}_{F}^{\mu}=\sum_{i=1}^{n}\sum_{j=1}^{i}\left(-1\right)^{j+1}D_{\mu_{j+1}}\cdots D_{\mu_{k}}Q^{\alpha}\left[D_{\mu_{1}}\cdots D_{\mu_{j-1}}\frac{\partial\mathcal{L}_{F}}{\partial\left(D_{\mu_{1}}\cdots D_{\mu_{j-1}}D_{\mu}D_{\mu_{j+1}}\cdots D_{\mu_{i}}\psi^{\alpha}\right)}\right].\label{eq:P_F}
\end{equation}

Combing Eqs.\,(\ref{eq:pr-v-l_a}), (\ref{eq:first-three-pr-v-l_a}),
(\ref{eq:last-three}) and (\ref{eq:prv-L_F+L}), equation (\ref{eq:infinitesimal-criterion-1})
is then transformed into
\begin{align}
 & \sum_{a}\int_{a_{1}}^{a_{2}}\left[q_{a}^{\mu}\boldsymbol{E}_{\boldsymbol{\chi}_{a}\mu}\left(\ell_{a}\right)+\xi^{\mu}D_{\mu}\ell_{a}+D_{\mu}P_{a}^{\mu}+\boldsymbol{E}_{\boldsymbol{\psi}}\left(\ell_{a}\right)\cdot\boldsymbol{Q}\right]d\tau_{a}\nonumber \\
 & +\xi^{\mu}D_{\mu}\mathcal{L}_{F}+D_{\mu}\mathbb{P}_{F}^{\mu}+\boldsymbol{E}_{\boldsymbol{\psi}}\left(\mathcal{L}_{F}\right)\cdot\boldsymbol{Q}+\mathcal{L}D_{\mu}\xi^{\mu}=0,\label{eq:41}
\end{align}
where we used 
\begin{align}
 & \sum_{a}\int_{a_{1}}^{a_{2}}\frac{d}{d\tau_{a}}\left[\ell_{a}\zeta_{a}+q_{a}^{\mu}\frac{\partial\ell_{a}}{\partial\dot{\chi}_{a}^{\mu}}+\frac{q_{a}^{\mu}}{c^{2}}\dot{\chi}_{a\mu}\left(\ell_{a}-\frac{\partial\ell_{a}}{\partial\dot{\chi}_{a}^{\nu}}\dot{\chi}_{a}^{\nu}\right)\right]d\tau_{a}\nonumber \\
 & =\left[\ell_{a}\zeta_{a}+q_{a}^{\mu}\frac{\partial\ell_{a}}{\partial\dot{\chi}_{a}^{\mu}}+\frac{q_{a}^{\mu}}{c^{2}}\dot{\chi}_{a\mu}\left(\ell_{a}-\frac{\partial\ell_{a}}{\partial\dot{\chi}_{a}^{\nu}}\dot{\chi}_{a}^{\nu}\right)\right]_{a_{1}}^{a_{2}}=0.
\end{align}
Suppose $\boldsymbol{\xi}$ and $\boldsymbol{Q}$ are independent
of $\tau_{a}$, equation (\ref{eq:41}) becomes 
\begin{align}
 & D_{\mu}\left[\mathcal{L}\xi^{\mu}+\mathbb{Q}^{\mu}\right]+\sum_{a}\int_{a_{1}}^{a_{2}}\left[q_{a}^{\mu}\boldsymbol{E}_{\boldsymbol{\chi}_{a}\mu}\left(\ell_{a}\right)\right]d\tau_{a}+\boldsymbol{E}_{\boldsymbol{\psi}}\left(\mathcal{L}\right)\cdot\boldsymbol{Q}=0,\label{eq:43}
\end{align}
where 
\begin{equation}
\mathbb{Q}^{\mu}=\sum_{i=1}^{n}\sum_{j=1}^{i}\left(-1\right)^{j+1}D_{\mu_{j+1}}\cdots D_{\mu_{k}}Q^{\alpha}\left[D_{\mu_{1}}\cdots D_{\mu_{j-1}}\frac{\partial\mathcal{L}}{\partial\left(D_{\mu_{1}}\cdots D_{\mu_{j-1}}D_{\mu}D_{\mu_{j+1}}\cdots D_{\mu_{i}}\psi^{\alpha}\right)}\right].\label{eq:P}
\end{equation}
Using the EL equation (\ref{eq:Psi-EL-equation}) of the field $\boldsymbol{\psi}$,
the last term in the Eq.\,(\ref{eq:43}) vanishes. However, due to
the the weak ELB equation (\ref{eq:weak-ELB}), the second term (\ref{eq:43})
is not zero. If the characteristic $\boldsymbol{q}_{a}$ are independent
of $\boldsymbol{\chi}$ and $\boldsymbol{\psi}$, this term can be
written as a divergence form, i.e.,
\begin{equation}
\sum_{a}\int_{a_{1}}^{a_{2}}\left[q_{a}^{\mu}\boldsymbol{E}_{\boldsymbol{\chi}_{a}\mu}\left(\ell_{a}\right)\right]d\tau_{a}=D_{\nu}\mathbb{S}^{\nu}\label{eq:45}
\end{equation}
where 
\begin{equation}
\mathbb{S}^{\nu}=\sum_{a}\int_{a_{1}}^{a_{2}}d\tau_{a}\left\{ q_{a}^{\mu}\left[-\ell_{a}\eta_{\:\mu}^{\nu}+\dot{\chi}_{a}^{\nu}\left[\frac{\partial\ell_{a}}{\partial\dot{\chi}_{a}^{\mu}}+\frac{1}{c^{2}}\dot{\chi}_{a\mu}\left(\ell_{a}-\dot{\chi}_{a}^{\sigma}\frac{\partial\ell_{a}}{\partial\dot{\chi}_{a}^{\sigma}}\right)\right]\right]\right\} \label{eq:S}
\end{equation}
Substituting Eq.\,(\ref{eq:45}) into Eq.\,(\ref{eq:43}), we finally
arrive at the geometric conservation law 
\begin{equation}
D_{\mu}\left[\mathcal{L}\xi^{\mu}+\mathbb{Q}^{\mu}+\mathbb{S}^{\mu}\right]=0.\label{eq:general-conservation-law}
\end{equation}

\section{gauge-symmetric energy-momentum conservation laws for high-order
electromagnetic systems coupled with charged particles \label{sec:gauge-invariant-energy-momentum}}

We now apply the general theory to high-order electromagnetic system
coupled with charged particles. The Lagrangian $L_{a}$ for this system
can be generally written as a gauge symmetric form as
\begin{align}
 & L_{a}=L_{a}\left(\chi_{a}^{\mu},A_{\mu},F_{\mu\nu},DF_{\mu\nu},\cdots,D^{\left(n\right)}F_{\mu\nu}\right),\label{eq:EM-La}\\
 & \mathcal{L}_{F}=\mathcal{L}_{F}\left(\chi^{\mu},F_{\mu\nu},DF_{\mu\nu},\cdots,D^{\left(n\right)}F_{\mu\nu}\right).\label{eq:EM-L-F}
\end{align}
where $D$ is the space-time derivative operator, i.e., $D=\left(\left(1/c\right)\partial/\partial t,\boldsymbol{\nabla}\right)$,
$A_{\mu}$ as the field $\boldsymbol{\psi}$ is the 4-potential and
$F_{\mu\nu}$ is the Faraday tensor defined by
\begin{equation}
F_{\mu\nu}=\partial_{\mu}A_{\nu}-\partial_{\nu}A_{\mu}.\label{eq:Faraday-tensor}
\end{equation}
Using Eq.\,(\ref{eq:Psi-EL-equation}), we can obtain the equation
of motion for electromagnetic field, 
\begin{equation}
E_{A}^{\mu}\left(\mathcal{L}\right)=0,\label{eq:Maxwell-Eq-1}
\end{equation}
where $E_{A}^{\mu}$ is the Euler operator for $A_{\mu}.$ To obtain
gauge-symmetric energy-momentum conservation laws, we transform (\ref{eq:Maxwell-Eq-1})
into
\begin{equation}
\partial_{\mu}\mathcal{D}^{\mu\nu}=\frac{4\pi}{c}J_{f}^{\nu}\label{eq:Maxwell-Eq-2}
\end{equation}
using the derivatives with respect to $F_{\mu\nu}$ (see Ref.\,\citep{Fan2021}),
where 
\begin{equation}
\mathcal{D}^{\mu\nu}=-8\pi E_{F}^{\left[\mu\nu\right]}\left(\mathcal{L}\right)\label{eq:displacement-tensor-1}
\end{equation}
is regarded as the electric displacement tensor. Here, $E_{F}^{\mu\nu}$
is the Euler operator for the Faraday tensor, i.e.,
\begin{equation}
E_{F}^{\mu\nu}\coloneqq\frac{\partial}{\partial F_{\mu\nu}}+\sum_{i=1}^{n}\left(-1\right)^{i}D_{\mu_{1}}\cdots D_{\mu_{i}}\frac{\partial}{\partial\partial_{\mu_{1}}\cdots\partial_{\mu_{i}}F_{\mu\nu}},\label{eq:Euler-operator-F}
\end{equation}
 $J_{f}^{\nu}$ is the free 4-current which is defined by 
\begin{equation}
J_{f}^{\nu}\coloneqq-c\frac{\partial\mathcal{L}}{\partial A_{\nu}}=-\int_{a_{1}}^{a_{2}}\frac{\partial\ell_{a}}{\partial A_{\nu}}ds_{a}.\label{eq:free-4-current}
\end{equation}
Here, $ds_{a}=cd\tau_{a}$ is the line element in the Minkowski space.
The superscript $[\mu\thinspace\nu]$ in Eq.\,(\ref{eq:displacement-tensor-1})
denotes the anti-symmetrization with respect of $\mu$ and $\nu$.
Suppose Eq.\,(\ref{eq:Maxwell-Eq-2}) is gauge symmetric and the
Lagrangian density obey the minimal coupling which can be written
as
\begin{equation}
\mathcal{L}=\sum_{a}\int_{a_{1}}^{a_{2}}\left[-\frac{q_{a}}{c}A_{\mu}\dot{\chi}_{a}^{\mu}\right]\delta_{a}d\tau_{a}+\mathrm{GSP}\left(\mathcal{L}\right)=-\frac{1}{c}A_{\mu}J^{\mu}+\mathrm{GSP}\left(\mathcal{L}\right),\label{eq:EM-Lagrangian-density}
\end{equation}
where $q_{a}$ is the charge of the $a$th particle and ``$\mathrm{GSP}\left(\mathcal{L}\right)$''
denotes the gauge-symmetric parts of the Lagrangian density. To get
the gauge-symmetric energy-momentum conservation laws, the prolongation
formula is also need reconstructed by using the derivatives with respect
to $F_{\mu\nu},$ that is 
\begin{align}
 & \text{pr}^{\left(n\right)}\boldsymbol{v}=\boldsymbol{v}+\sum_{a}\theta_{a1}^{\mu}\frac{\partial}{\partial\dot{\chi}_{a}^{\mu}}+\left[G_{\sigma\rho}+\xi^{\sigma}D_{\sigma}F_{\sigma\rho}\right]\frac{\partial}{\partial F_{\sigma\rho}}\nonumber \\
 & +\sum_{i=1}^{n}\left[D_{\mu_{1}}\cdots D_{\mu_{i}}G_{\sigma\rho}+\xi^{\sigma}D_{\sigma}D_{\mu_{1}}\cdots D_{\mu_{i}}F_{\sigma\rho}\right]\frac{\partial}{\partial\left(\partial_{\mu_{1}}\cdots\partial_{\mu_{i}}F_{\sigma\rho}\right)},\label{eq:prv-F_mu_nu}
\end{align}
where 
\begin{equation}
G_{\sigma\rho}=\partial_{\sigma}Q_{\rho}-\partial_{\rho}Q_{\sigma}=2\partial_{[\sigma}Q_{\rho]}.\label{eq:G}
\end{equation}
Finally, the conservation law (\ref{eq:general-conservation-law})
now read 
\begin{equation}
D_{\mu}\left\{ \mathcal{L}\xi^{\mu}-\frac{1}{4\pi}\mathcal{D}^{\mu\sigma}Q_{\sigma}+\mathbb{P}^{\mu}+\mathbb{S}^{\mu}\right\} =0,\label{eq:EM-general-conservation}
\end{equation}
where 
\begin{equation}
\mathbb{P}^{\mu}=\sum_{i=1}^{n}\sum_{j=1}^{i}\left(-1\right)^{j+1}\left(D_{\mu_{j+1}}\cdots D_{\mu_{i}}G_{\sigma\rho}\right)\left[D_{\mu_{1}}\cdots D_{\mu_{j-1}}\frac{\partial\mathcal{L}}{\partial\left(\partial_{\mu_{1}}\cdots\partial_{\mu_{j-1}}\partial_{\mu}\partial_{\mu_{j+1}}\cdots\partial_{\mu_{i}}F_{\sigma\rho}\right)}\right].\label{eq:EM-P}
\end{equation}

We now turn to discuss the space-time translation symmetry and energy-momentum
conservation law. Suppose the action of the system is invariant under
the space-time translation,
\begin{equation}
g_{\epsilon}\cdot\left(\tau_{a},\chi^{\mu},\chi_{a}^{\mu},A^{\mu}\right)=\left(\tilde{\tau}_{a},\tilde{\chi}^{\mu},\tilde{\chi}_{a}^{\mu},\tilde{A}^{\mu}\right)=\left(\tau_{a},\chi^{\mu}+\epsilon\chi_{0}^{\mu},\chi_{a}^{\mu}+\epsilon\chi_{0}^{\mu},A^{\mu}\right).\label{eq:spacetime-translation}
\end{equation}
The corresponding infinitesimal generator of the group transformation
(\ref{eq:spacetime-translation}) is
\begin{equation}
\boldsymbol{v}=\chi_{0}^{\mu}\frac{\partial}{\partial\chi^{\mu}}+\sum_{a}\chi_{0}^{\mu}\frac{\partial}{\partial\chi_{a}^{\mu}},\label{eq:space-infinitesimal-generator}
\end{equation}
where $\zeta_{a}=\phi^{\alpha}=0,\;\xi^{\mu}=\theta_{a}^{\mu}=\chi_{0}^{\mu}.$
The characteristics $\boldsymbol{Q}$, $\boldsymbol{q}_{a}$ and the
term $G_{\sigma\rho}$ are calculated as 
\begin{align}
 & q_{a}^{\mu}\equiv\chi_{0}^{\mu},\label{eq:EM-q_a}\\
 & Q^{\alpha}\equiv-\chi_{0}^{\nu}\partial_{\nu}A^{\alpha},\label{eq:EM-Q}\\
 & G_{\sigma\rho}=-\chi_{0}^{\nu}\partial_{\nu}F_{\sigma\rho}.\label{eq:EM-G}
\end{align}
It is clear that the second infinitesimal criterion (\ref{eq:infinitesimal-criterion-2})
is satisfied by $\dot{q}_{a}^{\mu}\equiv0$. Substituting Eqs.\,(\ref{eq:EM-q_a})-(\ref{eq:EM-G})
into Eq.\,(\ref{eq:EM-general-conservation}), we obtain the canonical
energy-momentum conservation law,
\begin{align}
 & D_{\mu}T_{N}^{\mu\nu}=0,\label{eq:energy-momentum-conservation}\\
 & T_{N}^{\mu\nu}=\mathcal{L}\eta^{\mu\nu}+\frac{1}{4\pi}\mathcal{D}^{\mu\sigma}\partial^{\nu}A_{\sigma}-\Sigma^{\mu\nu}+\Pi^{\mu\nu},\label{eq:energy-momentum-tensor}
\end{align}
where
\begin{align}
 & \Sigma^{\mu\nu}=\sum_{i=1}^{n}\sum_{j=1}^{i}\left(-1\right)^{j+1}\left(D_{\mu_{j+1}}\cdots D_{\mu_{i}}\partial^{\nu}F_{\sigma\rho}\right)\left[D_{\mu_{1}}\cdots D_{\mu_{j-1}}\frac{\partial\mathcal{L}}{\partial\left(\partial_{\mu_{1}}\cdots\partial_{\mu_{j-1}}\partial_{\mu}\partial_{\mu_{j+1}}\cdots\partial_{\mu_{i}}F_{\sigma\rho}\right)}\right],\label{eq:EM-Sigma}\\
 & \Pi^{\mu\nu}=\sum_{a}\int_{a_{1}}^{a_{2}}d\tau_{a}\left\{ -\ell_{a}\eta^{\mu\nu}+\dot{\chi}_{a}^{\mu}\left[\frac{\partial\ell_{a}}{\partial\dot{\chi}_{a\nu}}+\frac{1}{c^{2}}\dot{\chi}_{a}^{\nu}\left(\ell_{a}-\dot{\chi}_{a}^{\sigma}\frac{\partial\ell_{a}}{\partial\dot{\chi}_{a}^{\sigma}}\right)\right]\right\} .\label{eq:em-Pi}
\end{align}
Here, $T_{N}^{\mu\nu}$ is the canonical energy-momentum tensor which
is gauge dependent. We next ``improve'' $T_{N}^{\mu\nu}$ to a gauge-symmetric
form. We add the following identity 
\begin{equation}
D_{\mu}\left(D_{\sigma}\mathcal{F}^{\sigma\mu\nu}\right)=0,\quad\mathcal{F}^{\sigma\mu\nu}\equiv\frac{1}{4\pi}\mathcal{D}^{\sigma\mu}A^{\nu}\label{eq:F=00003DD*A}
\end{equation}
to Eq.\,(\ref{eq:energy-momentum-conservation}) to get the explicitly
gauge invariant conservation law
\begin{equation}
D_{\mu}T_{\text{GS}}^{\mu\nu}=0,
\end{equation}
where
\begin{equation}
T_{\text{GS}}^{\mu\nu}=\mathcal{L}\eta^{\mu\nu}+\frac{1}{c}J_{f}^{\mu}A^{\nu}+\Pi^{\mu\nu}+\frac{1}{4\pi}\mathcal{D}^{\mu\sigma}F_{\;\sigma}^{\nu}-\Sigma^{\mu\nu}\label{eq:T-inv}
\end{equation}
is the improved energy-momentum tensor, where we used Eq.\,(\ref{eq:Maxwell-Eq-2}).
We next prove that $T_{\text{GS}}^{\mu\nu}$ is gauge invariant. It
is sufficient to show that the first three terms in the right-hand
side of Eq.\,(\ref{eq:T-inv}) is gauge invariant. Substituting Eqs.\,(\ref{eq:EM-Lagrangian-density})
and (\ref{eq:em-Pi}) into Eq.\,(\ref{eq:T-inv}), these terms are
\begin{align}
 & \mathcal{L}\eta^{\mu\nu}+\frac{1}{c}J_{f}^{\mu}A^{\nu}+\Pi^{\mu\nu}\nonumber \\
 & =\mathcal{L}\eta^{\mu\nu}+\frac{1}{c}J_{f}^{\mu}A^{\nu}+\sum_{a}\int_{a_{1}}^{a_{2}}d\tau_{a}\left\{ -\ell_{a}\eta^{\mu\nu}+\dot{\chi}_{a}^{\mu}\left[\frac{\partial\ell_{a}}{\partial\dot{\chi}_{a\nu}}+\frac{1}{c^{2}}\dot{\chi}_{a}^{\nu}\left(\ell_{a}-\dot{\chi}_{a}^{\sigma}\frac{\partial\ell_{a}}{\partial\dot{\chi}_{a}^{\sigma}}\right)\right]\right\} \nonumber \\
 & =\mathcal{L}\eta^{\mu\nu}-\left(\sum_{a}\int_{a_{1}}^{a_{2}}d\ell_{a}\tau_{a}\right)\eta^{\mu\nu}+\frac{1}{c}J_{f}^{\mu}A^{\nu}\nonumber \\
 & +\sum_{a}\int_{a_{1}}^{a_{2}}d\tau_{a}\left\{ -\frac{q_{a}}{c}\dot{\chi}_{a}^{\mu}A^{\nu}+\dot{\chi}_{a}^{\mu}\text{GSP}\left(\frac{\partial\ell_{a}}{\partial\dot{\chi}_{a\nu}}\right)+\frac{1}{c^{2}}\dot{\chi}_{a}^{\mu}\dot{\chi}_{a}^{\nu}\left[\ell_{a}-\frac{q_{a}}{c}\dot{\chi}_{a}^{\sigma}A_{\sigma}-\dot{\chi}_{a}^{\sigma}\text{GSP}\left(\frac{\partial\ell_{a}}{\partial\dot{\chi}_{a}^{\sigma}}\right)\right]\right\} \nonumber \\
 & =\mathcal{L}_{F}\eta^{\mu\nu}+\frac{1}{c}J_{f}^{\mu}A^{\nu}-\frac{1}{c}\left(\sum_{a}\int_{a_{1}}^{a_{2}}q_{a}\dot{\chi}_{a}^{\mu}d\tau_{a}\right)A^{\nu}\nonumber \\
 & +\frac{1}{c^{2}}\sum_{a}\int_{a_{1}}^{a_{2}}d\tau_{a}\left\{ \dot{\chi}_{a}^{\mu}\text{GSP}\left(\frac{\partial\ell_{a}}{\partial\dot{\chi}_{a\nu}}\right)+\dot{\chi}_{a}^{\mu}\dot{\chi}_{a}^{\nu}\left[\text{GSP}\left(\ell_{a}\right)-\dot{\chi}_{a}^{\sigma}\text{GSP}\left(\frac{\partial\ell_{a}}{\partial\dot{\chi}_{a}^{\sigma}}\right)\right]\right\} \nonumber \\
 & =\mathcal{L}_{F}\eta^{\mu\nu}+\sum_{a}\int_{a_{1}}^{a_{2}}d\tau_{a}\left\{ \dot{\chi}_{a}^{\mu}\left[\text{GSP}\left(\ell_{a}\right)-\dot{\chi}_{a}^{\sigma}\text{GSP}\left(\frac{\partial\ell_{a}}{\partial\dot{\chi}_{a}^{\sigma}}\right)\right]\right\} ,\label{eq:Proof}
\end{align}
where Eq.\,(\ref{eq:free-4-current}) is used in the last step. Equation
(\ref{eq:Proof}) confirms that $T_{\text{inv}}^{\mu\nu}$ is gauge
invariant.

Lastly, we discuss the Podolsky system \citep{Podolsky1942} coupled
with charged particles, where the Lagrangian density of particles
and fields are respectively written as
\begin{align}
 & \ell_{\text{P}a}=-\left(m_{a}c^{2}+\frac{q_{a}}{c}A_{\mu}\dot{\chi}_{a}^{\mu}\right)\delta_{a}\label{eq:Po-L_a}\\
 & \mathcal{L}_{\text{PF}}=-\frac{1}{16\pi}F_{\sigma\rho}F^{\sigma\rho}-\frac{a^{2}}{8\pi}\partial_{\sigma}F^{\sigma\lambda}\partial^{\rho}F_{\rho\lambda}.\label{eq:Po-L_F}
\end{align}
Substituting Eq.\,(\ref{eq:Po-L_a}) into Eq.\,(\ref{eq:em-Pi})
to obtain $\Pi_{\text{P}}^{\mu\nu}$
\begin{equation}
\Pi_{\text{P}}^{\mu\nu}=\sum_{a}\int_{a_{1}}^{a_{2}}d\tau_{a}\left\{ \left[m_{a}c^{2}+\frac{q_{a}}{c}A_{\sigma}\dot{\chi}_{a}^{\sigma}\right]\delta_{a}\eta^{\mu\nu}-\left[m_{a}\dot{\chi}_{a}^{\mu}\dot{\chi}_{a}^{\nu}+\frac{q_{a}}{c}\dot{\chi}_{a}^{\mu}A^{\nu}\right]\delta_{a}\right\} .\label{eq:pO-Pi}
\end{equation}
$\Sigma^{\mu\nu}$ and $\mathcal{D}^{\sigma\mu}$ in Eq.\,(\ref{eq:T-inv})
has been shown in Ref.\,\citep{Fan2021}, we rewritten here as
\begin{align}
\Sigma_{\text{P}}^{\mu\nu} & =-\frac{a^{2}}{4\pi}\left(\partial^{\nu}F_{\;\rho}^{\mu}\right)\left(\partial_{\sigma}F^{\sigma\rho}\right),\label{eq:Po-Sigma}\\
\mathcal{D}_{\text{P }}^{\sigma\mu} & =-F^{\mu\sigma}+a^{2}\left(\partial^{\mu}\partial_{\lambda}F^{\lambda\sigma}-\partial^{\sigma}\partial_{\lambda}F^{\lambda\mu}\right).\label{eq:Po-dispacement tensor}
\end{align}
Using Eqs.\,\ref{eq:Po-L_a})-(\ref{eq:Po-dispacement tensor}),
the gauge invariant energy-momentum tensor is now become
\begin{align}
 & -T_{\text{inv}}^{\mu\nu}=\sum_{a}\int_{a_{1}}^{a_{2}}\left(m_{a}\dot{\chi}_{a}^{\mu}\dot{\chi}_{a}^{\nu}\delta_{a}\right)d\tau_{a}+\frac{1}{4\pi}\left[-F^{\mu\sigma}F_{\nu\sigma}+\frac{1}{4}\left(F_{\sigma\rho}F^{\sigma\rho}\right)\eta^{\mu\nu}\right]\nonumber \\
 & +\frac{a^{2}}{8\pi}\left(\partial_{\sigma}F^{\sigma\lambda}\partial^{\rho}F_{\rho\lambda}\right)\eta^{\mu\nu}+\frac{a^{2}}{4\pi}F_{\;\sigma}^{\nu}\left(\partial^{\mu}\partial_{\rho}F^{\rho\sigma}\right)-\frac{a^{2}}{4\pi}F_{\;\sigma}^{\nu}\left(\partial^{\sigma}\partial_{\rho}F^{\rho\mu}\right)-\frac{a^{2}}{4\pi}\left(\partial^{\nu}F_{\;\rho}^{\mu}\right)\left(\partial_{\sigma}F^{\sigma\rho}\right).\label{eq:Po-energy-momentum-conservation}
\end{align}

\section{conclusion and discussion}

In this work, we developed a general geometric (or manifestly covariant)
field theory for classical relativistic particle-field systems and
established the connections between general symmetries and local conservation
laws for the systems. To achieve this goal, we overcame two difficulties.

The first difficulty associated with the mass-shell constraint (see
Eq.\,(\ref{eq:mass-shell})). As a result, the standard Euler-Lagrange
(EL) equation is reconstructed by the Euler-Lagrange-Barut (ELB) equation
(see Eq.\,(\ref{eq:weak-ELB})). Besides, the use of proper time
parameter makes the Lagrangian density (\ref{eq:Lagrangian-density})
a function of the field $\boldsymbol{\psi}$ and also a functional
of the particle's world line, which directly lead to the standard
infinitesimal criterion a integro-differential equation rather than
a differential one (see Eq.\,(\ref{eq:infinitesimal-criterion-1})).
Furthermore, to satisfy the mass-shell condition, an extra criterion
(\ref{eq:infinitesimal-criterion-2}) is derived.

The second difficulty comes from the heterogeneous-manifolds that
the particles and fields reside on. The fields are defined on the
4D space-time, while each particle\textquoteright s word line is defined
only on the 1D parameter space. As a consequence, the standard Noether\textquoteright s
procedure for deriving local conservation laws from symmetries is
not applicable without modification. To overcome this difficulty,
we developed a weak version of the Euler-Lagrange-Barut (ELB) equation
for particles on the 4D space-time, which is rewritten here as
\begin{equation}
E_{\boldsymbol{\chi}_{a}\mu}\left(\ell_{a}\right)=D_{\nu}H_{\thinspace\mu}^{\nu},
\end{equation}
where the definition of $H_{\thinspace\mu}^{\nu}$ can be easily read
from Eq.\,(\ref{eq:weak-ELB}). This non-vanishing term $H_{\thinspace\mu}^{\nu}$
is emerged in the transformed infinitesimal criterion (\ref{eq:43})
as 
\begin{equation}
\sum_{a}\int_{a_{1}}^{a_{2}}\left[q_{a}^{\mu}D_{\nu}H_{\thinspace\mu}^{\nu}\right]d\tau_{a}.\label{eq:81}
\end{equation}
If the characteristic $q_{a}^{\mu}$ is independent of space-time
position $\boldsymbol{\chi}$ and the field $\boldsymbol{\psi}$,
the derivative operator $D_{\nu}$ can be moved out from the integral,
i.e., $\sum_{a}\int_{a_{1}}^{a_{2}}\left[q_{a}^{\mu}H_{\thinspace\mu}^{\nu}\right]d\tau_{a}=D_{\nu}\mathcal{J^{\nu}},$the
new current $\mathcal{J^{\nu}}\equiv\sum_{a}\int_{a_{1}}^{a_{2}}\left[q_{a}^{\mu}H_{\thinspace\mu}^{\nu}\right]d\tau_{a}$
is then induced.

Combing the weak ELB equation and infinitesimal criterion of the symmetry
conditions, the general geometric conservation laws is systematically
derived. Using general theory constructed here, we obtain the energy-momentum
conservation laws for high-order relativistic electromagnetic systems
by the space-time translation symmetry.

Interestingly, when the characteristic $q_{a}^{\mu}$ of the transformation
(\ref{eq:group-transformation}) is related with space-time position
$\boldsymbol{\chi}$ or the field $\boldsymbol{\psi}$, the derivative
operator $D_{\nu}$ cannot be moved out from the integral. As such,
equation (\ref{eq:81}) cannot be transformed into a divergence form,
and a conservation law cannot be given even the system admitting a
continuous symmetry. We have not find a suitable symmetry which won't
lead to a conservation law. Here, we recommend it as an open question
and it would be an exciting research project to shed more light on
that question.
\begin{acknowledgments}
P. Fan was supported by Shenzhen Clean Energy Research Institute and
National Natural Science Foundation of China (NSFC-12005141). Q. Chen
was supported by the National Natural Science Foundation of China
(NSFC-11805273). J. Xiao was supported by the National MC Energy R\&D
Program (2018YFE0304100), National Key Research and Development Program
(2016YFA0400600, 2016YFA0400601 and 2016YFA0400602), and the National
Natural Science Foundation of China (NSFC-11905220).
\end{acknowledgments}

\bibliographystyle{apsrev4-1}
\bibliography{CovariantConservationLaws}

\begin{thebibliography}{23}%
\makeatletter
\providecommand \@ifxundefined [1]{%
 \@ifx{#1\undefined}
}%
\providecommand \@ifnum [1]{%
 \ifnum #1\expandafter \@firstoftwo
 \else \expandafter \@secondoftwo
 \fi
}%
\providecommand \@ifx [1]{%
 \ifx #1\expandafter \@firstoftwo
 \else \expandafter \@secondoftwo
 \fi
}%
\providecommand \natexlab [1]{#1}%
\providecommand \enquote  [1]{``#1''}%
\providecommand \bibnamefont  [1]{#1}%
\providecommand \bibfnamefont [1]{#1}%
\providecommand \citenamefont [1]{#1}%
\providecommand \href@noop [0]{\@secondoftwo}%
\providecommand \href [0]{\begingroup \@sanitize@url \@href}%
\providecommand \@href[1]{\@@startlink{#1}\@@href}%
\providecommand \@@href[1]{\endgroup#1\@@endlink}%
\providecommand \@sanitize@url [0]{\catcode `\\12\catcode `\$12\catcode
  `\&12\catcode `\#12\catcode `\^12\catcode `\_12\catcode `\%12\relax}%
\providecommand \@@startlink[1]{}%
\providecommand \@@endlink[0]{}%
\providecommand \url  [0]{\begingroup\@sanitize@url \@url }%
\providecommand \@url [1]{\endgroup\@href {#1}{\urlprefix }}%
\providecommand \urlprefix  [0]{URL }%
\providecommand \Eprint [0]{\href }%
\providecommand \doibase [0]{http://dx.doi.org/}%
\providecommand \selectlanguage [0]{\@gobble}%
\providecommand \bibinfo  [0]{\@secondoftwo}%
\providecommand \bibfield  [0]{\@secondoftwo}%
\providecommand \translation [1]{[#1]}%
\providecommand \BibitemOpen [0]{}%
\providecommand \bibitemStop [0]{}%
\providecommand \bibitemNoStop [0]{.\EOS\space}%
\providecommand \EOS [0]{\spacefactor3000\relax}%
\providecommand \BibitemShut  [1]{\csname bibitem#1\endcsname}%
\let\auto@bib@innerbib\@empty
\bibitem [{\citenamefont {Ma}\ and\ \citenamefont {Wang}(1995)}]{Ma1995}%
  \BibitemOpen
  \bibfield  {author} {\bibinfo {author} {\bibfnamefont {C.~Y.}\ \bibnamefont
  {Ma}}\ and\ \bibinfo {author} {\bibfnamefont {D.~Y.}\ \bibnamefont {Wang}},\
  }\href {\doibase 10.1063/1.871350} {\bibfield  {journal} {\bibinfo  {journal}
  {Physics of Plasmas}\ }\textbf {\bibinfo {volume} {2}},\ \bibinfo {pages}
  {1361} (\bibinfo {year} {1995})}\BibitemShut {NoStop}%
\bibitem [{\citenamefont {Wei}\ and\ \citenamefont {Wu}(2021)}]{Wei2021}%
  \BibitemOpen
  \bibfield  {author} {\bibinfo {author} {\bibfnamefont {J.-J.}\ \bibnamefont
  {Wei}}\ and\ \bibinfo {author} {\bibfnamefont {X.-F.}\ \bibnamefont {Wu}},\
  }\href {\doibase 10.1007/s11467-021-1049-x} {\bibfield  {journal} {\bibinfo
  {journal} {Frontiers of Physics}\ }\textbf {\bibinfo {volume} {16}},\
  \bibinfo {pages} {44300} (\bibinfo {year} {2021})}\BibitemShut {NoStop}%
\bibitem [{\citenamefont {Beklemishev}\ and\ \citenamefont
  {Tessarotto}(2004)}]{Beklemishev2004}%
  \BibitemOpen
  \bibfield  {author} {\bibinfo {author} {\bibfnamefont {A.}~\bibnamefont
  {Beklemishev}}\ and\ \bibinfo {author} {\bibfnamefont {M.}~\bibnamefont
  {Tessarotto}},\ }\href {\doibase 10.1051/0004-6361:20034208} {\bibfield
  {journal} {\bibinfo  {journal} {Astronomy {\&} Astrophysics}\ }\textbf
  {\bibinfo {volume} {428}},\ \bibinfo {pages} {1} (\bibinfo {year}
  {2004})}\BibitemShut {NoStop}%
\bibitem [{\citenamefont {Xie}\ and\ \citenamefont {Du}(2007)}]{Xie2007}%
  \BibitemOpen
  \bibfield  {author} {\bibinfo {author} {\bibfnamefont {B.-s.}\ \bibnamefont
  {Xie}}\ and\ \bibinfo {author} {\bibfnamefont {S.-c.}\ \bibnamefont {Du}},\
  }\href {\doibase 10.1007/s11467-007-0036-1} {\bibfield  {journal} {\bibinfo
  {journal} {Frontiers of Physics in China}\ }\textbf {\bibinfo {volume} {2}},\
  \bibinfo {pages} {178} (\bibinfo {year} {2007})}\BibitemShut {NoStop}%
\bibitem [{\citenamefont {Littlejohn}(1984)}]{Littlejohn1984}%
  \BibitemOpen
  \bibfield  {author} {\bibinfo {author} {\bibfnamefont {R.~G.}\ \bibnamefont
  {Littlejohn}},\ }\href {\doibase 10.1063/1.864688} {\bibfield  {journal}
  {\bibinfo  {journal} {Physics of Fluids}\ }\textbf {\bibinfo {volume} {27}},\
  \bibinfo {pages} {976} (\bibinfo {year} {1984})}\BibitemShut {NoStop}%
\bibitem [{\citenamefont {Beklemishev}\ and\ \citenamefont
  {Tessarotto}(1999)}]{Beklemishev2012}%
  \BibitemOpen
  \bibfield  {author} {\bibinfo {author} {\bibfnamefont {A.}~\bibnamefont
  {Beklemishev}}\ and\ \bibinfo {author} {\bibfnamefont {M.}~\bibnamefont
  {Tessarotto}},\ }\href {\doibase 10.1063/1.873736} {\bibfield  {journal}
  {\bibinfo  {journal} {Physics of Plasmas}\ }\textbf {\bibinfo {volume} {6}},\
  \bibinfo {pages} {4487} (\bibinfo {year} {1999})}\BibitemShut {NoStop}%
\bibitem [{\citenamefont {Boghosian}(1987)}]{Boghosian2003b}%
  \BibitemOpen
  \bibfield  {author} {\bibinfo {author} {\bibfnamefont {B.~M.}\ \bibnamefont
  {Boghosian}},\ }\href {http://arxiv.org/abs/physics/0307148} {\bibfield
  {journal} {\bibinfo  {journal} {arxiv: physics/0307148}\ } (\bibinfo {year}
  {1987})}\BibitemShut {NoStop}%
\bibitem [{\citenamefont {Similon}(1985)}]{Similon1985}%
  \BibitemOpen
  \bibfield  {author} {\bibinfo {author} {\bibfnamefont {P.~L.}\ \bibnamefont
  {Similon}},\ }\href {\doibase 10.1016/0375-9601(85)90456-6} {\bibfield
  {journal} {\bibinfo  {journal} {Physics Letters A}\ }\textbf {\bibinfo
  {volume} {112}},\ \bibinfo {pages} {33} (\bibinfo {year} {1985})}\BibitemShut
  {NoStop}%
\bibitem [{\citenamefont {Brizard}\ and\ \citenamefont
  {Chan}(1999)}]{Brizard2011a}%
  \BibitemOpen
  \bibfield  {author} {\bibinfo {author} {\bibfnamefont {A.~J.}\ \bibnamefont
  {Brizard}}\ and\ \bibinfo {author} {\bibfnamefont {A.~A.}\ \bibnamefont
  {Chan}},\ }\href {\doibase 10.1063/1.873742} {\bibfield  {journal} {\bibinfo
  {journal} {Physics of Plasmas}\ }\textbf {\bibinfo {volume} {6}},\ \bibinfo
  {pages} {4548} (\bibinfo {year} {1999})}\BibitemShut {NoStop}%
\bibitem [{\citenamefont {Noether}(1918)}]{Noether1918}%
  \BibitemOpen
  \bibfield  {author} {\bibinfo {author} {\bibfnamefont {E.}~\bibnamefont
  {Noether}},\ }\href@noop {} {\bibfield  {journal} {\bibinfo  {journal}
  {Nachr. K{\"{o}}nig. Gesell. Wiss G{\"{o}}ttingen, Math. -Phys. Kl.}\
  }\textbf {\bibinfo {volume} {235}} (\bibinfo {year} {1918})},\ \bibinfo
  {note} {also available in English at Transport Theory and Statistical Physics
  1, 186-207 (1971) and arXiv: physics/0503066v3}\BibitemShut {NoStop}%
\bibitem [{\citenamefont {Peskin}\ and\ \citenamefont
  {Schroeder}(2018)}]{Peskin2018}%
  \BibitemOpen
  \bibfield  {author} {\bibinfo {author} {\bibfnamefont {M.~E.}\ \bibnamefont
  {Peskin}}\ and\ \bibinfo {author} {\bibfnamefont {D.~V.}\ \bibnamefont
  {Schroeder}},\ }\href@noop {} {\emph {\bibinfo {title} {{An Introduction to
  Quantum Field Theory}}}}\ (\bibinfo  {publisher} {CRS Press},\ \bibinfo
  {address} {New York},\ \bibinfo {year} {2018})\BibitemShut {NoStop}%
\bibitem [{\citenamefont {Landau}\ and\ \citenamefont
  {Lifshitz}(1971)}]{Landau1971}%
  \BibitemOpen
  \bibfield  {author} {\bibinfo {author} {\bibfnamefont {L.}~\bibnamefont
  {Landau}}\ and\ \bibinfo {author} {\bibfnamefont {E.}~\bibnamefont
  {Lifshitz}},\ }\href@noop {} {\emph {\bibinfo {title} {{The Classical Theory
  of Fields}}}},\ \bibinfo {edition} {4th}\ ed.\ (\bibinfo  {publisher}
  {Butterworth-Heinemann},\ \bibinfo {address} {Oxford},\ \bibinfo {year}
  {1971})\BibitemShut {NoStop}%
\bibitem [{\citenamefont {Fan}\ \emph {et~al.}(2018)\citenamefont {Fan},
  \citenamefont {Qin}, \citenamefont {Liu}, \citenamefont {Xiang},\ and\
  \citenamefont {Yu}}]{Fan2018}%
  \BibitemOpen
  \bibfield  {author} {\bibinfo {author} {\bibfnamefont {P.}~\bibnamefont
  {Fan}}, \bibinfo {author} {\bibfnamefont {H.}~\bibnamefont {Qin}}, \bibinfo
  {author} {\bibfnamefont {J.}~\bibnamefont {Liu}}, \bibinfo {author}
  {\bibfnamefont {N.}~\bibnamefont {Xiang}}, \ and\ \bibinfo {author}
  {\bibfnamefont {Z.}~\bibnamefont {Yu}},\ }\href {\doibase
  10.1007/s11467-018-0793-z} {\bibfield  {journal} {\bibinfo  {journal}
  {Frontiers of Physics}\ }\textbf {\bibinfo {volume} {13}},\ \bibinfo {pages}
  {135203} (\bibinfo {year} {2018})}\BibitemShut {NoStop}%
\bibitem [{\citenamefont {Qin}\ \emph {et~al.}(2014)\citenamefont {Qin},
  \citenamefont {Burby},\ and\ \citenamefont {Davidson}}]{Qin2014b}%
  \BibitemOpen
  \bibfield  {author} {\bibinfo {author} {\bibfnamefont {H.}~\bibnamefont
  {Qin}}, \bibinfo {author} {\bibfnamefont {J.~W.}\ \bibnamefont {Burby}}, \
  and\ \bibinfo {author} {\bibfnamefont {R.~C.}\ \bibnamefont {Davidson}},\
  }\href {\doibase 10.1103/PhysRevE.90.043102} {\bibfield  {journal} {\bibinfo
  {journal} {Physical Review E}\ }\textbf {\bibinfo {volume} {90}},\ \bibinfo
  {pages} {043102} (\bibinfo {year} {2014})}\BibitemShut {NoStop}%
\bibitem [{\citenamefont {Fan}\ \emph {et~al.}(2019)\citenamefont {Fan},
  \citenamefont {Qin}, \citenamefont {Xiao},\ and\ \citenamefont
  {Xiang}}]{Fan2019b}%
  \BibitemOpen
  \bibfield  {author} {\bibinfo {author} {\bibfnamefont {P.}~\bibnamefont
  {Fan}}, \bibinfo {author} {\bibfnamefont {H.}~\bibnamefont {Qin}}, \bibinfo
  {author} {\bibfnamefont {J.}~\bibnamefont {Xiao}}, \ and\ \bibinfo {author}
  {\bibfnamefont {N.}~\bibnamefont {Xiang}},\ }\href {\doibase
  10.1063/1.5092131} {\bibfield  {journal} {\bibinfo  {journal} {Physics of
  Plasmas}\ }\textbf {\bibinfo {volume} {26}},\ \bibinfo {pages} {062115}
  (\bibinfo {year} {2019})}\BibitemShut {NoStop}%
\bibitem [{\citenamefont {Fan}\ \emph {et~al.}(2020)\citenamefont {Fan},
  \citenamefont {Qin},\ and\ \citenamefont {Xiao}}]{Fan2020}%
  \BibitemOpen
  \bibfield  {author} {\bibinfo {author} {\bibfnamefont {P.}~\bibnamefont
  {Fan}}, \bibinfo {author} {\bibfnamefont {H.}~\bibnamefont {Qin}}, \ and\
  \bibinfo {author} {\bibfnamefont {J.}~\bibnamefont {Xiao}},\ }\href
  {http://arxiv.org/abs/2006.11039} {\bibfield  {journal} {\bibinfo  {journal}
  {arXiv: 2006.11039}\ } (\bibinfo {year} {2020})}\BibitemShut {NoStop}%
\bibitem [{\citenamefont {Barut}(1964)}]{Barut1964}%
  \BibitemOpen
  \bibfield  {author} {\bibinfo {author} {\bibfnamefont {A.~O.}\ \bibnamefont
  {Barut}},\ }\enquote {\bibinfo {title} {Electrodynamics and classical theory
  of fields {\&} particles},}\ \ (\bibinfo  {publisher} {Dover Publications,
  INC},\ \bibinfo {address} {New York},\ \bibinfo {year} {1964})\
  Chap.~\bibinfo {chapter} {II}, pp.\ \bibinfo {pages} {65--67}\BibitemShut
  {NoStop}%
\bibitem [{\citenamefont {Brehme}(1971)}]{Brehme1971}%
  \BibitemOpen
  \bibfield  {author} {\bibinfo {author} {\bibfnamefont {R.~W.}\ \bibnamefont
  {Brehme}},\ }\href {\doibase 10.1119/1.1986121} {\bibfield  {journal}
  {\bibinfo  {journal} {American Journal of Physics}\ }\textbf {\bibinfo
  {volume} {39}},\ \bibinfo {pages} {275} (\bibinfo {year} {1971})}\BibitemShut
  {NoStop}%
\bibitem [{\citenamefont {Qin}\ \emph {et~al.}(2007)\citenamefont {Qin},
  \citenamefont {Cohen}, \citenamefont {Nevins},\ and\ \citenamefont
  {Xu}}]{Qin2007a}%
  \BibitemOpen
  \bibfield  {author} {\bibinfo {author} {\bibfnamefont {H.}~\bibnamefont
  {Qin}}, \bibinfo {author} {\bibfnamefont {R.~H.}\ \bibnamefont {Cohen}},
  \bibinfo {author} {\bibfnamefont {W.~M.}\ \bibnamefont {Nevins}}, \ and\
  \bibinfo {author} {\bibfnamefont {X.~Q.}\ \bibnamefont {Xu}},\ }\href
  {\doibase 10.1063/1.2472596} {\bibfield  {journal} {\bibinfo  {journal}
  {Physics of Plasmas}\ }\textbf {\bibinfo {volume} {14}},\ \bibinfo {pages}
  {056110} (\bibinfo {year} {2007})}\BibitemShut {NoStop}%
\bibitem [{\citenamefont {Podolsky}(1942)}]{Podolsky1942}%
  \BibitemOpen
  \bibfield  {author} {\bibinfo {author} {\bibfnamefont {B.}~\bibnamefont
  {Podolsky}},\ }\href {\doibase 10.1103/PhysRev.62.68} {\bibfield  {journal}
  {\bibinfo  {journal} {Physical Review}\ }\textbf {\bibinfo {volume} {62}},\
  \bibinfo {pages} {68} (\bibinfo {year} {1942})}\BibitemShut {NoStop}%
\bibitem [{\citenamefont {Bopp}(1940)}]{Bopp1940}%
  \BibitemOpen
  \bibfield  {author} {\bibinfo {author} {\bibfnamefont {F.}~\bibnamefont
  {Bopp}},\ }\href {\doibase 10.1002/andp.19404300504} {\bibfield  {journal}
  {\bibinfo  {journal} {Annalen der Physik}\ }\textbf {\bibinfo {volume}
  {430}},\ \bibinfo {pages} {345} (\bibinfo {year} {1940})}\BibitemShut
  {NoStop}%
\bibitem [{\citenamefont {Olver}(1993)}]{Olver1988}%
  \BibitemOpen
  \bibfield  {author} {\bibinfo {author} {\bibfnamefont {P.~J.}\ \bibnamefont
  {Olver}},\ }\href@noop {} {\emph {\bibinfo {title} {{Applications of Lie
  Groups to Differential Equations}}}},\ \bibinfo {edition} {2nd}\ ed.\
  (\bibinfo  {publisher} {Springer-Verlag},\ \bibinfo {address} {New York},\
  \bibinfo {year} {1993})\BibitemShut {NoStop}%
\bibitem [{\citenamefont {Fan}\ \emph {et~al.}(2021)\citenamefont {Fan},
  \citenamefont {Xiao},\ and\ \citenamefont {Qin}}]{Fan2021}%
  \BibitemOpen
  \bibfield  {author} {\bibinfo {author} {\bibfnamefont {P.}~\bibnamefont
  {Fan}}, \bibinfo {author} {\bibfnamefont {J.}~\bibnamefont {Xiao}}, \ and\
  \bibinfo {author} {\bibfnamefont {H.}~\bibnamefont {Qin}},\ }\href
  {http://arxiv.org/abs/2103.14241} {\bibfield  {journal} {\bibinfo  {journal}
  {arxiv: 2103.14241}\ } (\bibinfo {year} {2021})}\BibitemShut {NoStop}%
\end{thebibliography}%

\end{document}